\documentclass{ws-rv9x6}
\usepackage{ws-rv-van}   

\def\laq{\raise 0.4ex\hbox{$<$}\kern -0.8em\lower 0.62ex\hbox{$\sim$}}
\def\gaq{\raise 0.4ex\hbox{$>$}\kern -0.7em\lower 0.62ex\hbox{$\sim$}}

\def\beq{\begin{equation}}
\def\eeq{\end{equation}}
\def\bea{\begin{eqnarray}}
\def\eea{\end{eqnarray}}

\def \pa {\partial}
\def \ra {\rightarrow}

\def \Da {\Delta}
\def \b {\beta}
\def \a {\alpha}

\def \Ga {\Gamma}
\def \ga {\gamma}

\def \da {\delta}
\def \ep {\epsilon}
\def \r {\rho}
\def \om {\omega}

\begin{document}

\begin{flushright}
Preprint BA-TH/803-20\\
\end{flushright}

\vspace{1cm}

\centerline{
{\large{
{\bf Gravity at Finite Temperature, Equivalence Principle,}}}}
\vspace{0.1cm}
\centerline{
{\large{
{\bf and Local Lorentz Invariance}}}}

\vspace{0.5cm}

\author[M. Gasperini]{M. Gasperini\footnote{E-mail address: maurizio.gasperini@ba.infn.it}
}

\address{Dipartimento di Fisica, Universit\`a di Bari,\\
Via G. Amendola 173, 70126 Bari, Italy, \\
and Istituto Nazionale di Fisica Nucleare, Sezione di Bari, Italy,\\
}

\vspace{1cm}

\centerline{Abstract}
\vspace{0.4cm}

\begin{abstract}
In this Chapter we illustrate the close connection between the violation of the weak equivalence principle typical of gravitational interactions at finite temperature, and similar violations induced by a breaking of the local Lorentz symmetry. We also discuss the physical implications of the effective repulsive forces possibly arising in such a generalized gravitational context, by considering, for an illustrative purpose, a quasi-Riemannian model of gravity with rotational symmetry as the local gauge group in tangent space.
\end{abstract}

\vspace{1.5 cm}

\centerline{ 
--------------------------------------------}

\vspace{0.5 cm}

\centerline{ 
To appear in  the book} 
\vspace{0.4cm}
\centerline{
 {\em ``Breakdown of the Einstein's Equivalence Principle''}} 
 \vspace{0.1cm}
\centerline{ed. by A. G. Lebed (World Scientific, 2021)}

\markboth{}{} 

\newpage

\body

\tableofcontents

\section{Introduction}
\label{sec1}
\renewcommand{\theequation}{1.\arabic{equation}}
\setcounter{equation}{0}

A breakdown of the Einstein's equivalence principle, which is the main subject of this book, is expected to occur also in the context of gravity at finite temperature, at the level of both classical/macroscopic and microscopic/quantum field interactions: in both cases there are indeed deviations from the standard, geodesic time-evolution and the locally-inertial type of motion. There are, however, important differences between the two cases.

In the case of classical test bodies one finds thermal geodesic deviations which depend on the mass  and total energy of the body\cite{1}, and which can be described by an effective geometry of the standard ``metric" type but with broken local Lorentz symmetry. The geodesic deviations at the quantum/microscopic level, on the contrary, are found to correspond to an effective {\em locally hyperbolic} type of free motion\cite{2} (characterized by a constant, nonvanishing, tangent-space acceleration) and require, for their classical description, a Lorentz invariant but ``metric-affine" (or Weyl) generalized geometrical structure.

In this Chapter we will concentrate on the first type of temperature-dependent effects, and we will show that the corresponding non-geodesic motion of massive, point-like bodies is just a particular case of more general equations of motion predicted by  matter-gravity interactions which are not locally Lorentz invariant, but only locally $SO(3)$-invariant\cite{3,4,5}. This suggests that an efficient classical description of  macroscopic gravity at finite temperature may be successfully implemented in the context of an effective geometric model different from General Relativity, and in which the local gauge symmetry of the $SO(3,1)$ group is broken also in the action describing the free gravitational dynamics. 

A possible example of such a gravitational theory is provided by the so-called ``quasi-Riemannian" models of gravity\cite{6,6a,6b,7}, and in particular by that class of models with local rotational symmetry in tangent space\cite{3,8,9}. Such an unconventional geometric structure is motivated by the fact that, at finite temperature, the flat tangent manifold describing the Minkowski vacuum has to be replaced by a tangent {\em thermal bath} at finite, nonvanishing temperature\cite{2}, which breaks the general $SO(3,1)$ invariance but preserves the local $SO(3)$ symmetry in the preferred  rest frame of the heat bath.

A gravitational model of this type may have interesting cosmological applications. In fact, the violation of the equivalence principle due to the breaking of the local Lorentz symmetry may be associated with the presence of effective repulsive interactions\cite{4,7,10,11}, whose consequences are relevant for both inflationary models\cite{9} and bouncing models preventing the initial singularity\cite{3,10}. This anticipates, in particular, scenarios very similar to the ones arising in a string cosmology context (see e.g. Refs. [\refcite{12,13,14,15}]).

In this Chapter we will present a short review of the above results obtained in previous papers and illustrating the close connection between gravity at finite temperature, gravitational interactions with broken local Lorentz symmetry, and violation of the weak equivalence principle. We will take the opportunity of clarifying  some technical details, not explicitly mentioned in the previous literature. 
The Chapter is organized as follows.

We will start in \sref{sec2} with the mass-dependent deviations from geodesic motion for free-falling test particles at finite temperature, and we shall discuss, in \sref{sec3}, the more general form of such deviations for gravitational interactions with broken local Lorentz symmetry. A simple, general covariant but locally only $SO(3)$-invariant model of gravity, and its possible cosmological consequences, will be briefly discussed in \sref{sec4}. A few concluding remarks will be finally presented in \sref{sec5}.

%
%

\section{Non-geodesic Motion of Test Particles at Finite Temperature}
\label{sec2}
\renewcommand{\theequation}{2.\arabic{equation}}
\setcounter{equation}{0}

Let us start by recalling that at finite temperature inertial and gravitational masses are in general different\cite{16,17}, as they correspond, from a thermodynamical point of view, to the low momentum limit of free energy and internal energy, respectively\cite{18}.

Considering in particular a charged particle of proper mass $m_0$, in thermal equilibrium with a photon heat bath at a temperature $T \ll m_0$, and in the absence of gravity, one finds that its total free energy $E$ can be written (to lowest order in $T^2$) as\cite{16,17}
\bea
&&
E(T)= \left[ m_0^2 + p^2 +{2\over 3} \a \pi T^2 \right]^{1/2} \equiv 
\left[ m^2(T) + p^2 \right]^{1/2} , 
\nonumber \\ && ~~
m(T) = \left[ m_0^2 + {2\over 3} \a \pi T^2 \right]^{1/2} \simeq m_0\left(1+ {1\over 3} \a T^2 \right),
\label{21}
\eea
where $\a$ is the fine structure constant and $m_0$  the (renormalized) rest mass at $T=0$. We are using units in which $\hbar$, $c$ and the Boltzmann constant $k_B$ are set equal to one.

On the other hand, according to the results of a detailed finite-temperature calculation performed in the weak field limit and in the rest frame of the eath bath\cite{16,17}, it turns out that the effective energy-momentum $\theta^{\mu\nu}$, representing the contribution of the same particle to the ``right-hand side" of the gravitational Einstein equations, can be expressed (again, to lowest order in $T^2$) as follows\cite{1}:
\beq
\theta^{\mu \nu}= T^{\mu\nu}-{2\over 3} \a \pi {T^2\over E^2(T)} V^\mu_4 V^\nu_4 T^{44}.
\label{22}
\eeq
Here $T^{\mu\nu}$ is the standard, minimally coupled to gravity, particle stress tensor (with temperature-dependent mass term), $T^{44}$ is the corresponding energy-density component locally evaluated in the flat tangent-space limit, and $E(T)$ is the (temperature-dependent) energy of Eq. (\ref{21}). Finally, $V^\mu_a$ is the so-called vierbein (or tetrad) field, connecting the world metric $g^{\mu\nu}$ of the given Riemann manifold to the flat Minkowski metric $\eta^{ab}$ of the local tangent space, in such a way that $g^{\mu\nu}= V^\mu_a V^\nu_b \eta^{ab}$. It follows that
\beq
T^{44}= T^{\a\b} V_\a^4 V_\b^4,
\label{23}
\eeq
and that the energy $E$ of Eq. (\ref{21}), representing the time-like component of the particle four-momentum at finite temperature in the locally flat space-time limit, can be related to the components of the ``curved" (i.e. generally covariant) momentum $p^\mu$ by:
\beq
E\equiv m \dot x^4 \equiv p^4 = p^\mu V_\mu^4 = m \dot x^\mu V_\mu^4
\label{24}
\eeq
(a dot denotes differentiation with respect to the proper time $\tau$).

It is appropriate, at this point, to clearly specify the adopted conventions. We will use Latin letters to denote flat (also called {\em anholonomic}) tangent space indices, and Greek letters to denote general-covariant ({\em holonomic}) world indices. Also, the index $``4"$ will always refer to the time-like coordinate $x^4$ of tangent space, while the index $``0"$ to the time-like coordinate $x^0 \equiv t$ of the curved Riemann manifold.

Given the above definitions and conventions, it is clear that the energy-momentum (\ref{22}) correctly transforms according to the tensor representation of the general diffeomorphism group $x^\mu \ra x^{\prime \mu}(x)$ acting on the coordinates of the Riemann manifold, but is not a scalar object under the action of the Lorentz symmetry group in the local Minkowski tangent space. However, the energy-momentum (\ref{22}) is manifestly compatible with a local rotational symmetry: the invariance under local transformations of the $O(3)$ group acting on the flat (Latin) indices. This is consistent with the presence in the local tangent space of a preferred frame at rest with the heat bath, which is indeed the frame where the explicit form of the effective gravitational source (\ref{22}) has been computed, and where the thermal radiation is isotropically distributed.

Let us also assume, for the moment, that any {\em direct} modification of the free geometric dynamics due to the temperature is absent (or negligible), so that the ``left-hand side" of the Einstein equations keeps unchanged, and the effective gravitational equations at finite temperature take the form $G^{\mu\nu}= 8 \pi G \,\theta^{\mu\nu}$, where $G^{\mu\nu}$ is the usual Einstein tensor and $G$ the Newton constant. The contracted Bianchi identity $\nabla_\nu G^{\mu\nu}=0$, where $\nabla_\nu$ is the Riemann covariant derivative, thus implies the generalized conservation equation $\nabla_\nu \theta^{\mu\nu}=0$, which for the effective energy-momentum tensor (\ref{22}) can be written explicitly as follows:
\bea
&&
\pa_\nu(\sqrt{-g} \,T^{\mu\nu})-  {2\over 3} \a \pi\,\pa_\nu \left( \sqrt{-g} \,{T^2\over E^2} V^\mu_4 V^\nu_4 T^{\a\b} V_\a^4 V_\b^4\right)+
\nonumber \\ &&~~
+ \sqrt{-g}\, \Ga_{\nu \a}\,^\mu \left(T^{\a\nu} - {2\over 3} \a \pi
{T^2\over E^2}  V^\a_4 V^\nu_4T^{\b\rho} V_\b^4 V_\rho^4\right)=0.
\label{25}
\eea

We should now recall that, given a test body and the covariant conservation of its energy-momentum tensor, the corresponding equation of motion can be obtained by applying the so-called Papapetrou procedure\cite{19}: namely, by integrating the conservation equation over an infinitely extended space-like hypersurface $\Sigma$  intersecting the ``world-tube" of the body at a given time $t=$ const, and by expanding the gravitational field variables in power series around the world-line $x^\mu(t)$ of its center of mass. One obtains, in this way, a ``multipole" expansion of the equation of motion including, at any given order, the gravitational coupling to all corresponding (dipole, quadrupole, etc) internal momenta.

We are interested, in this paper, in the case of a structureless, point-like test body. We can then neglect the contribution of all the internal momenta, and describe the test body with a delta-function distribution of its energy-momentum density, defined (see e.g. Ref. [\refcite{20}]) by
\beq
T^{\mu\nu}(x')= {1\over \sqrt{-g}} \da^3(x'-x(t))\,{p^\mu p^\nu\over p^0}
\equiv {m\over \sqrt{-g}} \da^3(x'-x(t))\,{\dot x^\mu \dot x^\nu\over \dot x^0},
\label{26}
\eeq
where $p^\mu=m\dot x^\mu$ and $p^0=m \dot x^0= m dt/d\tau$. In such a case the volume integration over the space-like hypersurface $\Sigma$, namely $\int_\Sigma d^3x' \sqrt{-g} \nabla_\nu \theta^{\mu\nu}$, 
become trivial, and by applying the Gauss theorem to eliminate the integral of spatial divergences (there is no flux of $\theta^{\mu\nu}$ at spatial infinity), we obtain, from Eq. (\ref{25}):
\bea
{dp^\mu\over d\tau}+\Ga_{\nu\a}\,^\mu {p^\a p^\nu\over p^0} &-&
{2 \a \pi\over 3}{d\over dt} \left( {T^2\over E^2}
V^\mu_4 V^0_4 {p^\a p^\b \over p^0} V_\a^4 V_\b^4\right)
\nonumber \\ &-& 
{2 \a \pi\over 3}{T^2\over E^2}\,\Ga_{\nu\a}\,^\mu
V^\a_4 V^\nu_4 {p^\b p^\r \over p^0} V_\b^4 V_\r^4=0.
\label{27}
\eea

Finally, let us express the time derivatives in terms of the proper time parameter $\tau$, and multiply the above equation by $m^{-1} dt/d\tau= m^{-1} \dot x^0= p^0/m^2$. By recalling that $p^\mu= m\dot x^\mu$, and by using for $E$ the definition (\ref{24}), we obtain
\beq 
\ddot x^\mu +\Ga_{\nu\a}\,^\mu \dot x^\a \dot x^\nu -
{2 \a \pi\over 3}{T^2\over m^2}{d \over d\tau} \left(V^\mu_4 V^0_4 {m\over p^0}\right)
-{2 \a \pi\over 3}{T^2\over m^2}\Ga_{\nu\a}\,^\mu V^\a_4 V^\nu_4 =0.
\label{28}
\eeq
The last two terms describe the mass-dependent deviations from geodesic motion induced by the thermal corrections, to lowest order in $T^2/m^2$. A simple application of this equation, illustrating the non-universality of free-fall at finite temperature, will be presented in the next subsection.


\subsection{Example: radial motion in the Schwarzschild field}
\label{sec21}

Let us consider the radial trajectory of test particle in the Schwarzschild geometry produced by a central source of mass $M$ and described, in polar coordinates $x^\mu=(t,r,\theta, \phi)$ by the diagonal metric
\beq
g_{\mu\nu} dx^\mu dx^\nu= e^\nu dt^2- e^{-\nu} dr^2 - r^2( d\theta^2 +\sin^2 \theta d\phi^2), ~~~ e^\nu= 1 -2GM/r.
\label{29}
\eeq
The vierbein field is also diagonal, with 
\beq
V_\mu^4=\da_\mu^4 e^{\nu/2},~~~~~~~~~~~~~~
V^\mu_4=\da^\mu_4 e^{-\nu/2},
\label{210}
\eeq
and the relevant components of the Christoffel connection, for a radial trajectory with $\dot \theta=0$, $\dot \phi=0$, are given by
\beq
\Ga_{01}\,^0= {\nu'\over 2}, ~~~~~~~ \Ga_{00}\,^1= {\nu'\over 2}e^{2\nu}, ~~~~~~~
\Ga_{11}\,^1=- {\nu'\over 2},
\label{211}
\eeq
where a prime denotes differentiation with respect to $r$. 

The radial motion in this gravitational field, according to Eq. (\ref{28}), is the described by the following two independent equations,
\beq
\ddot t+ \dot \nu \dot t=0, ~~~~~~~ \ddot r+{\nu'\over 2}\left( e^{2\nu} \dot t^2 - \dot r^2 - {2 \a \pi\over 3}{T^2\over m^2}e^\nu\right)=0,
\label{212}
\eeq
whose integration (with the condition of vanishing radial velocity at spatial infinity, $\dot r \ra 0$ for $r \ra \infty$) gives
\beq
\dot t= e^{-\nu}, ~~~~~~~~~ \dot r^2 =1 - e^\nu + {2 \a \pi\over 3}{T^2\over m^2}\nu e^\nu.
\label{213}
\eeq
By inserting this result into Eq. (\ref{212}) we can finally write the generalized expression for the radial acceleration of a test particle at finite temperature as follows:
\beq
\ddot r= -{GM\over r^2} \left\{ 1- {2 \a \pi\over 3}{T^2\over m^2}\left[1 + \ln \left(1-{2GM\over r}\right) \right]\right\}.
\label{214}
\eeq

This result describes, for $T>0$, a non-universal, mass-dependent deviation from geodesic motion. The temperature-dependent corrections controlling the breaking of the equivalence principle are very small, however. In the weak field limit, in which terms higher than linear in the gravitational potential $GM/r$ are neglected, we can estimate that the effective difference $\Da m$ between inertial and gravitational mass, for a particle of rest mass $m_0$, is given by
\beq
{|\Da m |\over m_0} \sim \a {T^2\over m_0^2}.
\label{215}
\eeq
For macroscopic masses and ordinary values of the temperature this effect is well outside the present experimental sensitivities (see e.g. the results of the recent MICROSCOPE  space mission\cite{21}). Let us notice, for instance, that at a temperature $T \sim 300$ Kelvin degrees, and for an electron mass $m_0 \sim 0.5$ MeV, we have $\a (T/m_0)^2 \sim 10^{-18}$.

\begin{figure}[t]
\centering
\includegraphics[width=9cm]{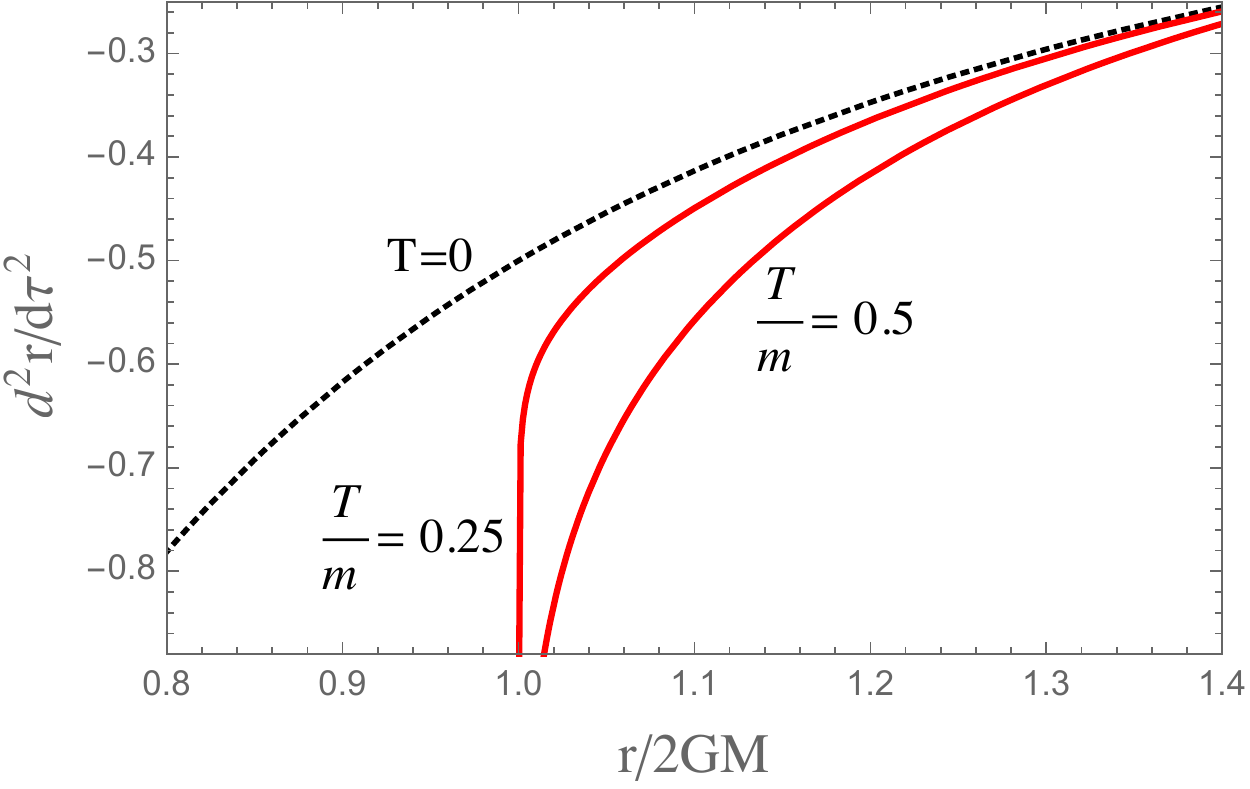}
\caption{The radial acceleration of Eq. (\ref{214}) is plotted without temperature corrections ($T=0$, black dashed curve) and with temperature corrections ($T>0$, red solid curves). At finite temperature  the attractive force tends to diverge at $r=2GM$ for any given value of the ratio $T/m <1$.} 
\label{f1}
\end{figure}

Assuming that the result (\ref{214}) for the radial acceleration keeps valid if extrapolated to the strong gravity regime (i.e. at very small values of the radial coordinate), it may be interesting to note that the deviations from the geodesic trajectory are still mass depend, but the gravitational attraction tends to diverge, for {\em any} given value of $m$, when approaching the Schwarzschild radius $r=2GM$ (as illustrated in Fig. \ref{f1}). 

Let us stress, however, that the result (\ref{214}) is only valid in the limit $T/m \ll1$. At higher temperatures, higher-order corrections to the particle trajectory (and possibly also to the effective space-time geometry, see \sref{sec4}), are needed.


\section{Non-geodesic Motion for Locally Lorentz-Noninvariant Matter-Gravity Interactions}
\label{sec3}
\renewcommand{\theequation}{3.\arabic{equation}}
\setcounter{equation}{0}

The energy-momentum tensor (\ref{22}), modified by the thermal corrections, is only a particular example of matter distribution coupled to gravity with general-covariant but not locally Lorentz-invariant interactions.

More generally, assuming that the gravitational coupling to matter is only $SO(3)$-invariant in the local tangent space, we can decompose the standard energy-momentum $T^{\mu\nu}$ of the matter sources into its tangent space components $T^{ij}$, $T^{i4}$ and $T^{44}$ transforming, respectively, as a tensor, a vector, and a scalar under the local $SO(3)$ group (conventions: Latin indices $i,j,k, \dots$ run from 1 to 3). If the local Lorentz symmetry is broken, those different components may contribute with different coupling strength to the gravitational equations\cite{3,4,5}, thus producing an effective source of gravity described by a modified energy-momentum tensor $\theta^{\mu\nu}= T^{\mu\nu} + \Da T^{\mu\nu}$.

For the purpose of this paper we can conveniently (and equivalently) work with the $SO(3)$-scalar variables $T^{44}$, $T^{4 \nu}$, $T^{\mu4}$, so as to express the modified gravitational source in general-covariant form, and in terms of the minimally coupled matter stress tensor $T^{\a\b}$, as follows\cite{3,4,5}:
\bea
\!\!\!\!\!\!
\theta^{\mu\nu} &=& T^{\mu\nu}  + a_1 \,V^\mu_4 V^\nu_4 T^{44}+ a_2 \,V^\mu_4 T^{4 \nu}+ a_3\, V^\nu_4 T^{\mu 4}
\nonumber \\
&\equiv&
T^{\mu\nu}+a_1\, V^\mu_4 V^\nu_4 T^{\a\b}V_\a^4 V_\b^4+
a_2 \,V^\mu_4 T^{\a \nu}V_\a^4 + a_3\, V^\nu_4 T^{\mu \a}V_\a^4.
\label{31}
\eea
Here $a_1$, $a_2$, $a_3$ are dimensionless parameters governing the breaking of the local $SO(3,1)$ symmetry, which is restored in the limit $a_1=a_2=a_3=0$. 

It should be noted that the generalized tensor $\theta^{\mu\nu}$ is symmetric, $\theta^{\mu\nu}= \theta^{\nu\mu}$, provided that $T^{\mu\nu}$ is symmetric and $a_2=a_3$. Note, also, that the finite-temperature stress tensor given by Eq. (\ref{22}) can be exactly reproduced from Eq. (\ref{31}) by putting $a_2=a_3=0$ and $a_1= - (2 \a \pi/3) (T^2/E^2)$. In this Section we shall assume that the coefficients $a_i$  are constant parameters; it should be stressed, however, that they might acquire an intrinsic energy dependence in a different (and probably more realistic) model of Lorentz symmetry breaking, as suggested indeed by the finite-temperature scenario discussed in \sref{sec2}. 

Let us now assume, as in \sref{sec2}, that local Lorentz symmetry is broken only in the matter part of the action, in such a way that the generalized gravitational equations can be written as $G^{\mu\nu}= 8 \pi G \theta^{\mu\nu}$, and the contracted Bianchi identity implies the conservation equation $\nabla_\nu \theta^{\mu\nu}=0$. For consistency with the symmetry property of the Einstein tensor $G^{\mu\nu}$ we have to assume, of course, that $\theta^{\mu\nu}$ is also symmetric, which implies (as previously stressed)  $T^{\mu\nu}= T^{\nu\mu}$ and $a_2=a_3$. The conservation law of the energy-momentum tensor (\ref{31}) thus provides the condition
\bea
&&
~~~~~~~~~~~~~~~~~
\pa_\nu(\sqrt{-g} \,T^{\mu\nu})+ a_1\, \pa_\nu \left(\sqrt{-g} \, V^\mu_4 V^\nu_4 T^{\a\b}V_\a^4 V_\b^4\right)+
\nonumber \\ &&
~~~~~~~~~~~~~~~~~
+a_2 \,\pa_\nu \left(\sqrt{-g} \,V^\mu_4 T^{\a \nu}V_\a^4 + \sqrt{-g} \, V^\nu_4 T^{\mu \a}V_\a^4 \right)+
\nonumber \\ &&
 +\sqrt{-g} \,\Ga_{\nu \a}\,^\mu (T^{\a\nu} +
 a_1 V^\a_4 V^\nu_4 T^{\b\rho}V_\b^4 V_\rho^4 +
 a_2V^\a_4 T^{\b \nu}V_\b^4 +a_2V^\nu_4 T^{\a \b}V_\b^4)=0
\nonumber \\ &&
\label{32}
\eea

We shall follow the same procedure as in \sref{sec2}, by integrating the above equation over an infinitely extended spatial hypersurface $\Sigma$, by applying the Gauss theorem to eliminate the integral of the spatial divergences, and by assuming that $T^{\mu\nu}$ can be appropriately described by the delta-function distribution (\ref{26}). By multiplying the result by $m^{-1} dt/d\tau$ we finally obtain the following generalized equation of motion,
\bea
&&
\ddot x^\mu +\Ga_{\nu\a}\,^\mu \dot x^\a \dot x^\nu +
a_1{d\over d\tau} \left(V^\mu_4 V^0_4 \,{\dot x^a \dot x^\b\over \dot x^0} V_\a^4 V_\b^4 \right)+
\nonumber \\ &&
+a_2\, {d\over d\tau} \left[ \left(V^\mu_4+ V^0_4 \,{\dot x^\mu\over \dot x^0} \right) \dot x^\a V_\a^4\right]+
+a_1\, \Ga_{\nu\a}\,^\mu V^\a_4 V^\nu_4 \dot x^\b \dot x^\rho V_\b^4 V_\rho^4+
\nonumber \\ &&
+a_2\, \Ga_{\nu\a}\,^\mu \left(V^\a_4 \dot x^\nu + V^\nu_4 \dot x^\a \right) \dot x^\b V_\b^4=0,
\label{33}
\eea
which describes the non-geodesic trajectory of a point-like test body coupled to gravity in a way which preserves the local rotational symmetry, but breaks in general the local Lorentz invariance. The geodesic deviations are controlled by the Lorentz-breaking parameters $a_1$ and $a_2$.

In the following subsection we shall apply this result to discuss the possible effects on the radial acceleration of a free-falling test body in the static, spherically symmetric field of a central source.


\subsection{Example: repulsive forces and possible gravitational non-universality}
\label{sec31}

Let us consider, as in \sref{sec21}, a radial motion in the Schwarzschild geometry described by the metric (\ref{29}). The relevant components of the vierbein field and of the Christoffel connection are given by Eqs. (\ref{210}), (\ref{211}). By writing explicitly the $\mu=0$ and $\mu=1$ components of the equation of motion (\ref{33}) we obtain, respectively, the following two independent equations\cite{4,5}:
\bea
&&
(1+a_2+2a_2)\, \ddot t +(1+a_2) \,\dot \nu \dot t=0,
\label{34} \\
&&
(1+a_2) \,\ddot r -{\nu'\over 2} \dot r^2 + (1+a_1+2a_2)\,{\nu'\over 2}e^{2\nu} \dot t^2=0.
\label{35}
\eea
Assuming that $1+a_1+2a_2\not=0$ and $1+a_2 \not=0$ (we expect indeed that all Lorentz-breaking corrections are small, $|a_1| \ll1$, $|a_2| \ll1$), we can define the convenient parameters
\beq
\b= {1+ a_2\over 1+a_1 +2 a_2}, ~~~~~~~~~~~~
\ga ={1\over 1+a_2},
\label{36}
\eeq
and we can express the integration of Eqs. (\ref{34}), (\ref{35}) (with the initial condition of vanishing radial velocity at spatial infinity) as follows:
\beq
\dot t= e^{-\b \nu}, ~~~~~~~~~~~
\dot r^2 = {1\over \b(2-\ga-2\b)}\left[ e^{\ga \nu} - e^{2\nu(1-\b)} \right].
\label{37}
\eeq
Finally, by inserting this result into Eq. (\ref{35}), we find that the generalized radial acceleration of the test body is given by:
\beq
\ddot r = -{GM\over r^2}  {1\over \b(2-\ga-2\b)} \left[ 2 (1-\b) \left(1-{2GM\over r} \right)^{1-2\b} -\ga  \left(1-{2GM\over r} \right)^{\ga-1}\right].
\label{38}
\eeq

For $\b=\ga=1$ we recover the standard result $\ddot r= -GM/r^2$. Note that the above expression is different from the modified trajectory described by Eq. (\ref{214}), because at finite temperature the Lorentz-breaking terms are controlled by the energy-dependent (and thus time-dependent) coefficient $T^2/E^2$, which provides additional contributions to the geodesic deviations through its nonvanishing time derivative (see Eq. (\ref{27})).  In this Section, instead, we have assumed constant Lorentz-breaking parameters, $\dot a_1=0$, $\dot a_2=0$. 

It may be interesting to note however that, even with such a simplified model, and for an appropriate choice of the Lorentz-breaking parameters, we may expect the automatic appearance of repulsive gravitational interactions. This occurs (for instance) if $1/2 <\b <1$ and $\b>1- \ga/2$; and the last condition, incidentally, is automatically satisfied if the motion has to preserve causality in the spatial region $r>2GM$ (see Ref. [\refcite{4}] for a detailed discussion of the allowed numerical values of $\b$ and $\ga$ in order to avoid the presence of imaginary radial velocities and space-like four-velocity vectors).

\begin{figure}[t]
\centering
\includegraphics[width=9cm]{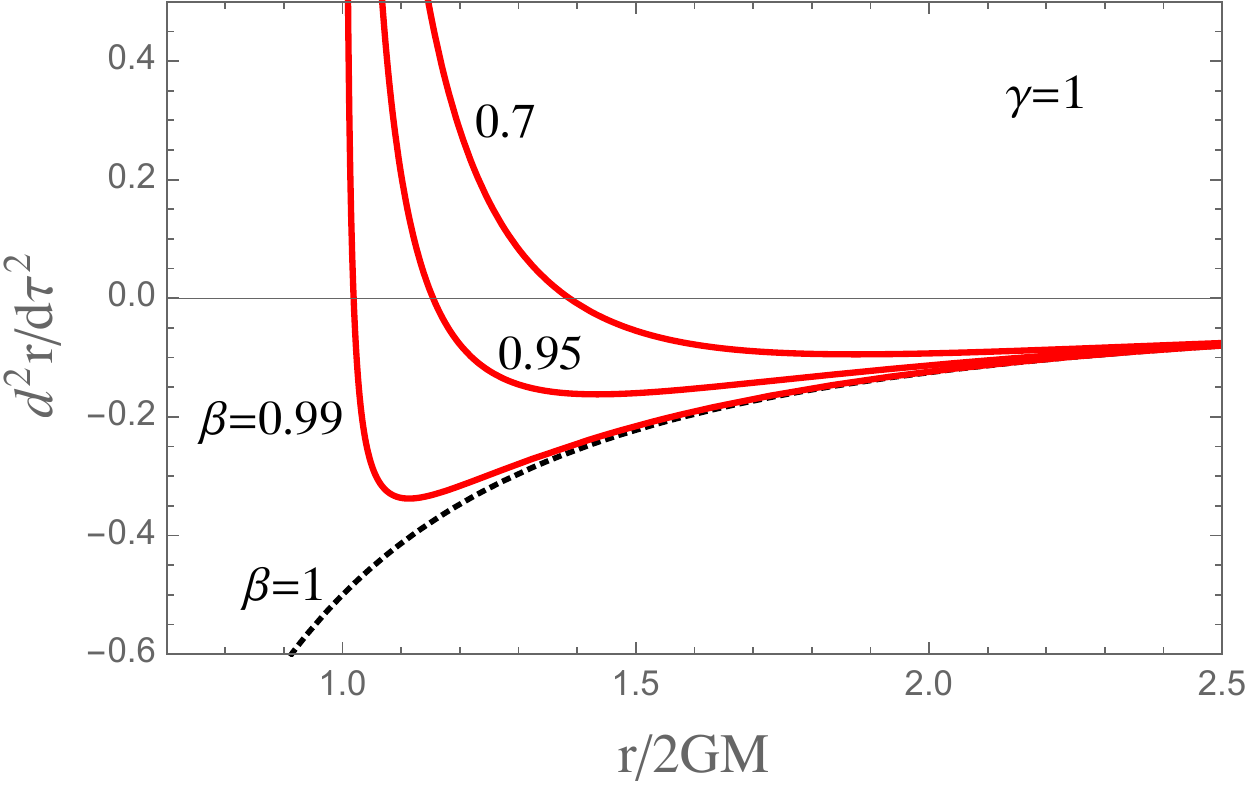}
\caption{The radial acceleration of Eq. (\ref{38}) is plotted for $\ga=1$ and for different values of $\b$. For $\b=1$ we have the standard result of general relativity  (black dashed curve). For $\b<1$ we have repulsive gravitational interactions which become dominant at small enough values of $r$, and diverge at $r=2GM$ (red solid curves with $\b=0.99$, $\b=0.95$ and $\b=0.7$).} 
\label{f2}
\end{figure}

The repulsive interactions, when present, become dominant at small enough radial distances, and tend to diverge in the limit $r \ra 2GM$, as illustrated in Fig. \ref{f2} for particular values of $\b$ and $\ga$ . In that case $\dot r \ra 0$ at $r=2GM$, and the interior of the Schwarzschild sphere becomes a ``classically impenetrable" region\cite{4} (an effect  similar to the one occurring in the context of 
Rosen bimetric theory of gravity\cite{22}). 

Let us stress that in the model we are considering the deviations from geodesic motion are triggered by the Lorentz-breaking parameters $a_1$, $a_2$ which are not necessarily mass dependent (unlike the finite-temperature corrections  discussed in \sref{sec2}). If so, the resulting free motion of a test particle in a given gravitational field is non-geodesic, but still ``universal".

In principle, however, the effective violation of the local Lorentz symmetry might be different for different types of particles, thus producing an effective non-universality of free-fall and of the gravitational coupling\cite{5,23,24}, which could be tested by applying the generalized equation of motion (\ref{38}).

For instance, let us assume (as a working hypothesis) that the local $SO(3,1)$ symmetry is broken for the gravitational interactions of baryons but not for those of leptons\cite{5}. This clearly produces a ``composition-dependent" violation of the equivalence principle, very similar, in practice, to that produced by the coupling of the so-called ``fifth force" to the baryon number\cite{25,26} (see also the recent discussion of Ref. [\refcite{28a}]). In such a case, if we have a macroscopic test body of mass $m_1$ containing $B_1$  baryons of mass $m_B$, and if we consider Eq. (\ref{38}) in the weak field limit (neglecting terms higher than linear in the gravitational potential $GM/r$), we find that the effective gravitational force acting on $m_1$ is given by
\beq
m_1 \ddot r_1= -{GM\over r^2} {m_1\over \b} = - {GM\over r^2} m_1 \left[ 1+ \left(a_1+a_2\over 1+a_2\right) {m_B\over m_1} B \right]
\label{39}
\eeq
(we have set $m_1= m_BB$).

Comparing the  accelerations $\ddot r_1$ and $\ddot r_2$ of two different test masses $m_1$ and $m_2$, in the Earth (or solar) gravitational field, we thus obtain
\beq
{\Da a \over g}= \left(a_1+a_2\over 1+a_2\right) \Da \left(B\over \mu\right),
\label{310}
\eeq
where $\Da a = \ddot r_1-\ddot r_2$, where $g= -GM/r^2$ is the local acceleration of gravity, and where $\Da (B/\mu)= (B_1/\mu_1)-(B_2/\mu_2)$, with $\mu=m/m_B$  the mass of the test body in units of baryonic mass. Hence, a different violation of the local Lorentz symmetry for baryons and leptons leads to a composition-dependent gravitational acceleration of macroscopic test bodies, which -- for constant values of $a_1$ and $a_2$ -- is strongly constrained by existing experimental results.  

According to the most recent tests of the equivalence principle\cite{21} we can impose, in fact, the upper limit $(\Da a /g)\, \laq \,10^{-15}$, for bodies with $\Da (B/\mu) \sim 10^{-3}$. This implies
\beq\left|a_1+a_2\over 1+a_2 \right| \,\laq \,10^{-12}
\label{311}
\eeq
(unless, of course, we consider more sophisticated models of Lorentz symmetry breaking where the constant parameters $a_1$, $a_2$ are replaced by position-dependent and/or energy-dependent variables).


\section{A Quasi-Riemannian Model of Gravity with Local $SO(3)$ Tangent-Space Symmetry}
\label{sec4}
\renewcommand{\theequation}{4.\arabic{equation}}
\setcounter{equation}{0}

We have shown, in the previous Sections, that a breaking of the local $SO(3,1)$ symmetry -- like that occurring at finite temperature -- leads to modify the  coupling of the test bodies to the background geometry. In such a context, it may be natural to expect  a modified dynamics also for the geometry itself: in particular, a dynamics described by gravitational equations which are not locally Lorentz-invariant but only $SO(3)$-invariant.

A simple way to formulate an effective theory of this type is to follow the scheme of the so-called ``quasi-Riemannian" models of gravity\cite{6,6a,6b,7}, and to choose, in particular the rotational group $SO(3)$ as the dynamical gauge symmetry of the flat space-time locally tangent to the (curved) world manifold\cite{3,4,5}. 

It is convenient, to this purpose, to construct the action working directly in the local tangent space, where the Lorentz connection $\om_\mu\,^{ab}$ can be decomposed into the $SO(3)$ connection $\om_\mu\,^{ij}$ and the $SO(3)$ vector  $\om_\mu\,^{i4}$ (let us recall that $i,j,k, \dots$ run from 1 to 3).  Using these variables, plus the local components of the vierbein field $V_\mu^i$, $V_\mu^4$ (transforming, respectively, as an $SO(3)$ vector and scalar field), we can then easily write a modified gravitational action which is generally covariant but, locally, only $SO(3)$ invariant.

It may be useful, also, to adopt the compact language of differential forms, and work with the connection one-form $\om^{ab} \equiv \om_\mu\,^{ab}dx^\mu$ and the anholonomic basis one-form $V^a \equiv V_\mu^a dx^\mu$. In this formalism the standard Einstein action can be written in terms of the curvature two-form $R^{ab}$ as
\bea
S_E&=& -\int d^4x \sqrt{-g} R \equiv {1\over 2} \int R^{ab} \wedge V^c \wedge V^d \ep_{abcd}, 
\nonumber \\ 
R^{ab}&=& d \om^{ab} + \om^a\,_c \wedge \om^{cd}.
\label{41} \eea
Conventions: the symbol $``d"$  denotes exterior derivative, and the wedge symbol $``\wedge"$ exterior product; finally, $\ep$ is the totally antisymmetric Levi-Civita symbol of the flat tangent space. 

We can now introduce the possible Lorentz breaking -- but $SO(3)$ preserving -- contributions, and write the generalized gravitational action in quasi-Riemannian form as follows,
\bea
S&=&{1\over 16 \pi G} \int \left[ {1\over 2} \int R^{ab} \wedge V^c \wedge V^d \ep_{abcd}
+ \left( b_1 \,\overline R^{ij}  \wedge V^k \wedge V^4 + \right. \right.
\nonumber \\  
&+&\left. \left.
 b_2\, \overline D \om^{i4}  \wedge V^j \wedge V^k + b_3\, \om^i\,_4 \wedge \om^{j4} \wedge  V^k \wedge V^4\right) \ep_{ijk4} + \cdots\right],
\label{42}
\eea
where we have explicitly introduced the $SO(3)$ curvature (or Yang-Mills) term $\overline R^{ij} $ and the $SO(3)$ covariant (exterior) derivative $\overline D$, defined by:
\beq
\overline R^{ij}= d \om^{ij} + \om^i\,_k\wedge \om^{kj}, ~~~~~~~~~~
\overline D \om^{i4}= d \om^{i4} + \om^i\,_k\wedge \om^{k4}.
\label{43}
\eeq
The dimensionless coefficients $b_i$ are constant parameters controlling the breaking of the local Lorentz symmetry, and the stand gravitational theory is recovered in the limit $b_i=0$. Finally, the dots denote the possible addition of other $SO(3)$-invariant contributions, that may be present or not depending on the chosen model of Lorentz-symmetry braking, as well as on the assumed type of geometry (e.g., with or without torsion, nonmetricity tensor, and so on). See Refs. 
[\refcite{3,7}]  for a general discussion.

By adding the action for the matter sources, and by varying the total action with respect to the $\om^{ab}$ and $V^a$ we then obtain, respectively, the explicit expression for the connection and the generalized form of the gravitational equations (see Refs. [\refcite{3,6a,7}]  for detailed computations). The final result for the modified Einstein equations can be written in general as
\beq
G^{\mu\nu} +\Da G^{\mu\nu}= 8 \pi G \theta^{\mu\nu} \equiv 8 \pi G \left(T^{\mu\nu} + \Da T^{\mu\nu} \right)
\label{44}
\eeq
where the right-hand side of these equations exactly corresponds to the generalized matter stress tensor $ \theta^{\mu\nu} $ of Eq. (\ref{31}). On the left-hand side we have the usual Einstein tensor, $G^{\mu\nu}= R^{\mu\nu} -Rg^{\mu\nu}/2$, plus the corrections $\Da G^{\mu\nu}$ induced by the breaking of the local Lorentz symmetry. 
  
It should be noted that $G^{\mu\nu}$ is a symmetric tensor, but $ \Da G^{\mu\nu} \not= \Da G^{\nu\mu}$, in general. Hence, there is no need of imposing on $\Da T^{\mu\nu}$ to be symmetric, and we may have $a_2 \not= a_3$ in Eq. (\ref{31}). 

Note also that the contracted Bianchi identity $\nabla_\nu G^{\mu\nu}=0$  leads to the conditions
\beq
\nabla_\nu T^{\mu\nu}= \nabla_\nu \left( {\Da G^{\mu\nu} \over 8 \pi G} - \Da T^{\mu\nu} \right),
\label{45}
\eeq
which implies, in general, deviations from the geodesic motion of free-falling test particles (as discussed in the previous sections). However, given that our modified gravitational equations depend on 6 (or more) parameters, $a_1, a_2, a_3, b_1, b_2, b_3, \dots$, it turns out that it is always possible in principle to preserve a geodesic type of motion ($\nabla_\nu T^{\mu\nu}=0$) by imposing as a constraint that the right-hand side of Eq. (\ref{45}) is identically vanishing. This constraint provides indeed four additional conditions which reduce the number of independent parameters for this class of models (see \sref{sec41} for an explicit example of this possibility).

For the illustrative purpose of this paper we shall concentrate  on a simple model of quasi-Riemannian gravity where the breaking of the local Lorentz symmetry leads to modified equations which can be written in terms of the Ricci tensor $R_\mu^\nu$ as follows: 
\bea
&& ~~~~~~~~~~~~~~~~~~~~~~
R_\mu\,^\nu + \Da R_\mu\,^\nu = 8 \pi G \left( \theta_\mu\,^\nu -{1\over 2} \da_\mu^\nu \theta \right), 
\label{46} \\ &&
\Da R_\mu\,^\nu =R_\mu\,^\nu+ c_1 \,R_{\mu 4}\,^{\nu 4}+ c_2 \,V_\mu^4 V_\nu^4 R_4\,^4+ c_3 \,V_\mu^4 R_4\,^\nu+ c_4 \,\om_{\mu\a 4} \om^{\nu \a 4} =
\nonumber \\ &&
= \left( c_1\, R_{\mu\a}\,^{\nu\b} + c_2\, V_\mu^4 V^\nu_4 R_\a\,^\b \right) V^\a_4 V_\b^4+ c_3\, V_\mu^4 R_\a\,^\nu V^\a_4 + c_4\, \om_{\mu\a\b} \om^{\nu \a \rho}V^\b_4 V_\r^4
\nonumber \\ &&
\label{47}
\eea
Here $\theta_\mu\,^\nu$ is given by Eq. (\ref{31}), $\theta = \theta_\a\,^\a$, $R_{\mu\a}\,^{\nu\b} $ is the usual Riemann tensor, and $c_1, \cdots , c_4$ are the constant Lorentz-breaking parameters. Finally, the tangent space connection $\om$ is fixed as usual by the so-called ``metricity postulate", 
\beq
\nabla_\mu V_\nu^a = \pa_\mu V_\nu^a + \om_\mu\,^a\,_b V_\nu^b - \Gamma_{\mu\nu}\,^\a V_\a^a =0.
\label{48}
\eeq
In the following subsection we shall apply the above equations to describe the cosmological geometry produced by a distribution of perfect-fluid matter sources.


\subsection{Example: cosmological applications with and without violation of the equivalence principle}
\label{sec41}

Let us consider the spatially homogeneous and isotropic Friedmann-Lemaitre-Robertson-Walker (FLRW) geometry, described in polar coordinates $x^\mu= ( t, r, \theta, \phi)$ by the metric
\beq
ds^2= g_{\mu\nu} dx^\mu dx^\nu= dt^2 - a^2(t) \left[ {dr^2\over 1-k r^2} + r^2 d\theta^2 + r^2 \sin^2 \theta d\phi^2 \right],
\label{49}
\eeq
where $t$ is the cosmic time, $a(t)$ the scale factor, and $k=0, \pm1$ the constant spatial curvature. The unperturbed energy-momentum $T^{\mu\nu}$ of the fluid sources, assumed at rest in the comoving frame, is given by the diagonal tensor
\beq
T_\mu\,^\nu= {\rm diag} \left( \r, -p, -p , -p \right),
\label{410}
\eeq
where the (time-dependent) energy density $\r$ and pressure $p$ are related by a barotropic equation of state, $p/\r=w=$ const.

For this geometry we simply have $V_\mu^4= \da_\mu^4$, $V^\mu_4=\da^\mu_4$, and the relevant components of the tangent space connection $\om$, fixed by Eq. (\ref{48}),  are given by $\om_\mu\,^\a\,_4= \Ga_{\mu 4}\,^\a = H ( \da_\mu^1 \da^\a_1+ \da_\mu^2 \da^\a_2+ \da_\mu^3\da^\a_3)$, where $H=\dot a/a$ (the dot denotes differentiation with respect to the cosmic time $t$). The generalized equations (\ref{46}) reduce, in this case, to the following two independent equations,
\bea
&&
-3 {\ddot a \over a} (1+c_2+c_3) = 4 \pi G \left[ \r (1+a_1+a_2+a_3) +3p \right],
\label{411}
\\ && \!\!\!\!\!\!\!\!\!\!\!\!\!
{\ddot a \over a}(1-c_1) +2 H^2 \left(1-{c_4\over 2} \right) +{2k\over a^2} = 4 \pi G 
 \left[ \r (1+a_1+a_2+a_3) -p \right],
\label{412}
\eea
obtained , respectively, from the $(0,0)$ and $(1,1)$ components of of Eq. (\ref{46}). In the limit $c_i=0$, $a_i=0$, we exactly recover the Einstein equations for the metric  (\ref{49}) and the matter distribution (\ref{410}).

Let us now consider two simple particular cases, describing ``minimal" (but interesting) modifications of the standard cosmological scenario.

The first one is based on the assumption that the Lorentz-breaking corrections lead to a new, modified cosmological dynamics which leaves unchanged, however, the form of the well-known Friedmann equation \cite{9}. Such a scenario can be obtained, in the context of our model, by the following choice of parameters:
\beq a_1=a_2=a_3=0, ~~~~~~~~  - c_1 = c_2+c_3 \not=0, ~~~~~~~~ c_4=0.
\label{413}
\eeq

With that choice, in fact, by eliminating $\ddot a/a$ from Eq. (\ref{411}) in terms of Eq. (\ref{412}), and using the identity $ \ddot a/a= \dot H +H^2$, we obtain that the two modified cosmological equations can be rewritten, respectively, as 
\bea
&&
H^2 +{k\over a^2}= {8 \pi G\over 3} \rho,
\label{414}
\\ &&
2 \dot H(1-c_1) + 3 H^2\left(1-{2\over 3} c_1 \right) +{k\over a^2} = -8 \pi G p,
\label{415}
\eea
and that their combination gives
\beq
\dot \r (1-c_1) +3 H(\r+p)= 2  c_1 H \r.
\label{416}
\eeq
We are thus left with an unchanged Friedmann equation (\ref{414}), but we have a corresponding non-trivial modification of the spatial Einstein equation (\ref{415}) and of the covariant evolution in time of the energy-momentum density, Eq. (\ref{416}) (which is no longer equivalent to the conservation law  $\nabla_\nu T^{\mu\nu}=0$). 

As discussed in Ref. [\refcite{9}], such a minimal, one-parameter-dependent violation of the local Lorentz symmetry may have interesting applications in a primordial cosmological context, where -- if the violation is strong enough -- it can produce accelerated (inflationary) expansion even in the absence of exotic sources with negative pressure (like, for instance, an effective cosmological constant). 

Consider in fact an early enough epoch when the Universe is still radiation dominated ($p=\r/3$), and the contribution of the spatial curvature to the cosmic dynamics is negligible, so that we can put $k=0$ in Eqs. (\ref{414})--(\ref{416}). A simple integration of those equations then gives
\beq
\r(t) \sim a ^{-(4-2c_1)/(1-c_1)},   ~~~~~~~~~~
a(t) \sim t^{(1-c_1)/(2-c_1)}.
\label{417}
\eeq
For $c_1>2$ this solution describes a phase of  accelerated expansion of ``power-law" type, with $\ddot a >0$ and $\dot H<0$. In the limit $c_1 \ra 2$ the solution describes a phase of exponential inflation, $a \ra \exp ( Ht)$, $\dot H \ra 0$ (with no need of introducing, to this purpose, an effective inflaton field assumed to be ``slow rolling" along some {\em ad hoc} inflaton potential). 

Let us now report the second example of modified cosmological dynamics where, in spite of the corrections due to the Lorentz-breaking terms, the covariant conservation of the standard energy-momentum tensor is preserved, $\nabla_\nu T^{\mu\nu}=0$, and the evolution in time of the matter sources is geodesic\cite{3,10}. This possibility corresponds to a model with the following values of the parameters:
\beq a_1=a_2=a_3=0, ~~~~~~~~  - c_1 = c_2+c_3 \not=0, ~~~~~~~~ c_4=2c_1 \not=0.
\label{418}
\eeq

In that case, by eliminating $\ddot a/a$ from Eq. (\ref{411}) in terms of Eq. (\ref{412}), we find that Eqs. (\ref{411}), (\ref{412}) can be rewritten  as 
\bea
&&
H^2(1-c_1) +{k\over a^2}= {8 \pi G\over 3} \rho,
\label{419}
\\ &&
2 \dot H(1-c_1) + 3 H^2\left(1-c_1 \right) +{k\over a^2} = -8 \pi G p,
\label{420}
\eea
and that their combination gives
\beq
\dot \r  +3 H(\r+p)= 0,
\label{421}
\eeq
which exactly corresponds to the standard conservation law of the energy-momentum (\ref{410}).

Note that in the absence of spatial curvature, $k=0$, this particular model of Lorentz symmetry breaking has no dynamical effects on the evolution of the cosmic geometry apart from a trivial renormalization of the coupling constant, $ G \ra G/(1-c_1)$. In that case, for $c_1>1$, we would find always repulsive gravitational interactions, a possibility which is clearly excluded by standard gravitational phenomenology.

Interestingly enough, however, if $k>0$ (and $c_1>1$), the repulsive interactions may become dominant in the limit of high enough energy density (i.e., during the very early cosmological phases), and allow non-singular ``bouncing" solutions to the cosmological equations even in the presence of conventional sources satisfying the strong energy condition\cite{3,10}. This is possible because, according  to the modified equations (\ref{419})--(\ref{421}), the condition which makes the singularity unavoidable (i.e. the condition of geodesic convergence $R_{\mu\nu} u^\mu u^\nu \geq 0$, where $u^\mu$ is a time-like vector field), and the strong energy condition, $(T_{\mu\nu}-g_{\mu\nu}T/2 \geq 0$, are no longer equivalent conditions. 

To give an explicit example let us consider a  radiation fluid with $p=\r/3$ and a cosmic geometry with $k=1/t_0^2>0$, where $t_0=$ const is a given parameter controlling the spatial curvature scale. 
From Eq. (\ref{421}) we obtain $\r=\r_0 a^{-4}$, where $\r_0$ is an integration constant, and the modified Friedmann equation (\ref{419}), with $c_1>1$, has the particular exact solution
\beq
a(t)= \left[ {8\pi G\over 3}\r_0 t_0^2 +{(t/t_0)^2\over |1-c_1|} \right]^{1/2},
\label{422}
\eeq
with the cosmic time $t$ ranging from $-\infty$ to $+\infty$. The associated Hubble parameter is given by
\beq
H= {t/t_0^2\over (t/t_0)^2 + |1-c_1| 8\pi G\r_0 t_0^2/3},
\label{423}
\eeq
and has no singularity in the whole range $-\infty \leq t \leq +\infty$. 

\begin{figure}[h]
\centering
\includegraphics[width=8.cm]{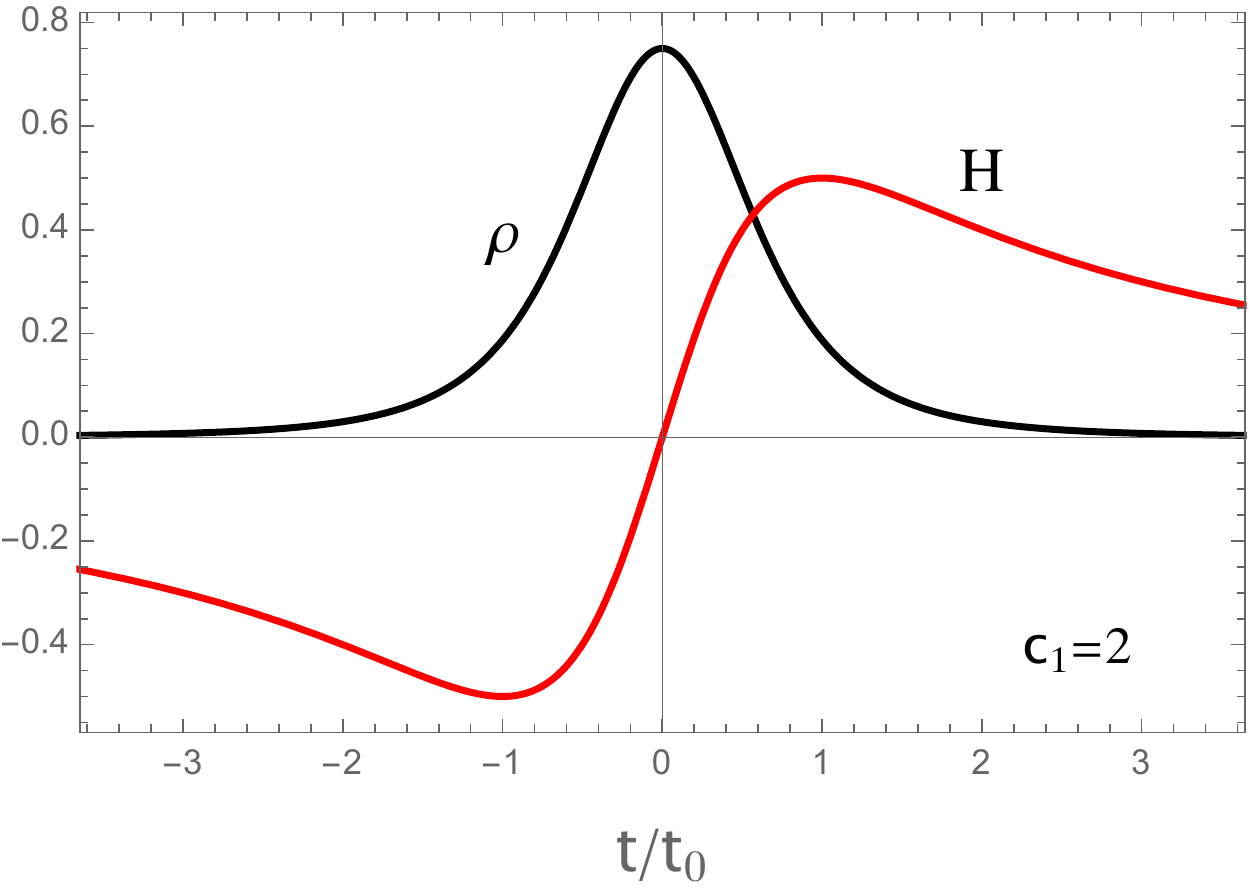}
\caption{Smoot evolution of the radiation energy density $\r= \r_0/a^4$ (black solid curve) and of the Hubble parameter $H$ (red solid curve) though the bouncing transition described by the solution (\ref{422}), (\ref{423}). The curves are plotted for $c_1=2$,  $8\pi G\r_0 t_0^2/3=1$, $\r_0=0.75$, and $H$ is expressed in units of $1/t_0$. } 
\label{f3}
\end{figure}


As illustrated in Fig. \ref{f3}, the above solution describes a continuous and regular bouncing transition between two complementary (or ``dual") cosmological phases\footnote{See Refs.  [\refcite{12,13,14,27,28}] for similar scenarios in a string cosmology context.}, defined, respectively, in the time ranges $t<0$ and $t>0$. 
The initial, asymptotically flat regime describes, for $ t<0$, a collapsing phase of decelerated contraction ($\dot a <0$, $\ddot a >0$), initially growing (in modulus) curvature scale ($\dot H<0$), and growing energy density ($\dot \r >0$). The density reaches the maximum value $\r_0$ at the epoch $t=0$, which marks a smooth transition towards the final regime characterized, for $t>0$ by accelerated expansion  ($\dot a >0$, $\ddot a >0$), eventually decreasing curvature scale ($\dot H<0$), and decreasing energy density ($\dot \r <0$).

Let us stress, finally, that the two examples of modified gravitational dynamics reported in this Section require, for an efficient application in a cosmological context, relatively large values of the Lorentz-breaking parameters ($|c_i | \sim 1$). Hence, they are expected to possibly describe a realistic scenario only at very early epochs, when the Universe approaches the Planck energy scale and/or the quantum gravity regime. We know indeed, from many experimental data, that at lower energies the local Lorentz symmetry of the gravitational interactions is very efficiently restored, and only very weak violations are possibly allowed.


\section{Conclusion}
\label{sec5}
\renewcommand{\theequation}{5.\arabic{equation}}
\setcounter{equation}{0}

The principle of equivalence is at the very ground of Einstein's theory of gravity. According to this principle the gravitational interaction can always be locally eliminated, and we can always locally reduce to the physics of the flat space-time, governed by the principle of Lorentz invariance. 

If the effective local Lorentz invariance is broken (for instance, due to the presence of a thermal bath at finite temperature), we can then expect violations of the equivalence principle. Conversely, violations of such a principle, and deviations from the geodesic motion of point-like test bodies, may correspond to a local symmetry different from the Lorentz one.

The Lorentz symmetry group, on the other hand, is the gauge group of the General Relativity theory of gravity (where the curvature tensor plays the role of the non-Abelian ``Yang-Mills field" of the local $SO(3,1)$ symmetry). If we adopt a different gauge group we can formulate a different gravitational theory, where the space-time geometry is still described in terms of Riemannian manifolds, but with a dynamics controlled by field equations different from Einstein's equations (see also Ref.  [\refcite{31}] for a recent discussion of the independence of general coordinate transformations and local Lorentz transformations). In this paper we have considered, in particular, a possibly modified gravitational dynamics based on the local gauge group  of the spatial rotations.  

In that case the principle of equivalence is not satisfied, in general, unless we impose appropriate constraints on the chosen model of Lorentz-symmetry breaking. In addition, the breaking can also produce gravitational interactions of repulsive type, which may have a relevant impact on  the primordial cosmological dynamics. 

On a macroscopic scale of energies and distances, however, we know that the possible violations of the local Lorentz symmetry and of the equivalence principle are both constrained by present experimental data to be extremely weak, and to produce only subdominant effects. Nevertheless, we believe that the possibility of such effects should be included when studying models and applications of the gravitational interaction at very high energies and in the quantum regime.


\section*{Acknowledgements}
This work is supported in part by INFN under the program TAsP ({\it Theoretical Astroparticle Physics}), and by the research grant number 2017W4HA7S ({\it NAT-NET: Neutrino and Astroparticle Theory Network}), under the program PRIN 2017, funded by the Italian Ministero dell'Universit\`a e della Ricerca (MUR). 



\end{document}